# Semi-supervised Medical Image Segmentation via Geometry-aware Consistency Training

Zihang Liu[1], Chunhui Zhao[1, *]

*[1] State Key Laboratory of Industrial Control Technology, College of Control Science and Engineering, Zhejiang University, Hangzhou, 310027, China.*

**Abstract:** The performance of supervised deep learning methods for medical image segmentation is often limited due to the scarcity of labeled data. As a promising research direction, semi-supervised learning addresses this dilemma by leveraging unlabeled data information to assist the learning process. In this paper, a novel geometry-aware semi-supervised learning framework is proposed for medical image segmentation. Considering that the hard-to-segment regions are mainly located around the object boundary, we introduce an auxiliary prediction task to learn the global geometric information. Based on the geometric constraint, the ambiguous boundary regions are emphasized through an exponentially weighted strategy during the model training to better exploit both labeled and unlabeled data. In addition, a dual-view network is designed to perform segmentation from different perspectives and reduce the prediction uncertainty. We illustrate the effectiveness of the proposed method on two public benchmark datasets, left atrium dataset and brain tumor dataset. Extensive experiments show that our method outperforms the state-of-the-art semi-supervised segmentation methods, demonstrating the effectiveness of our strategy for the challenging semi-supervised segmentation tasks.

*Corresponding author: Tel:86-571-87951879, Fax: 86-571-87951879, E-mail address: chhzhao@zju.edu.cn

## 1. Introduction

Precise and robust segmentation of organs or abnormal regions from medical images plays an essential role in clinical applications. The accurate segmentation results enable the quantitative analysis of anatomical structures, which provides useful basis for clinicians to diagnose or make risk assessment of relevant diseases [1]. Recent years have witnessed the remarkable progress of deep learning algorithms in medical image segmentation [2]-[5] The segmentation accuracy has been greatly improved accordingly. Nevertheless, previous methods mainly followed the fully supervised setting, which adopted an encoder-decoder network architecture (e.g., U-Net [2] for 2D image segmentation and V-Net [6] for 3D image segmentation) and formulated the problem as a pixel-wise classification task. As a major property of supervised learning, the advanced performance heavily relies on sufficient labeled data. Inadequate training data usually leads to over-fitting and sub-optimal segmentation results. However, it is expensive and laborious to delineate reliable annotations in medical domain. The scarcity of the labeled data motivates many annotation-efficient studies such as semi-supervised learning [7], [8], weakly supervised learning [9]-[11] and unsupervised domain adaptation [12]-[14]. Since it is clinically practical to obtain a small set of expert-examined labeled images and a large number of unlabeled images, we focus on semi-supervised segmentation in this work.

Semi-supervised segmentation learns from a limited amount of labeled data and sufficient unlabeled data, aiming to promote the learning process by leveraging unlabeled data information. As a promising research direction, it has been widely studied during the past few years. Existing methods can be roughly divided into two categories: self-training methods [15] and consistency-based methods [16].

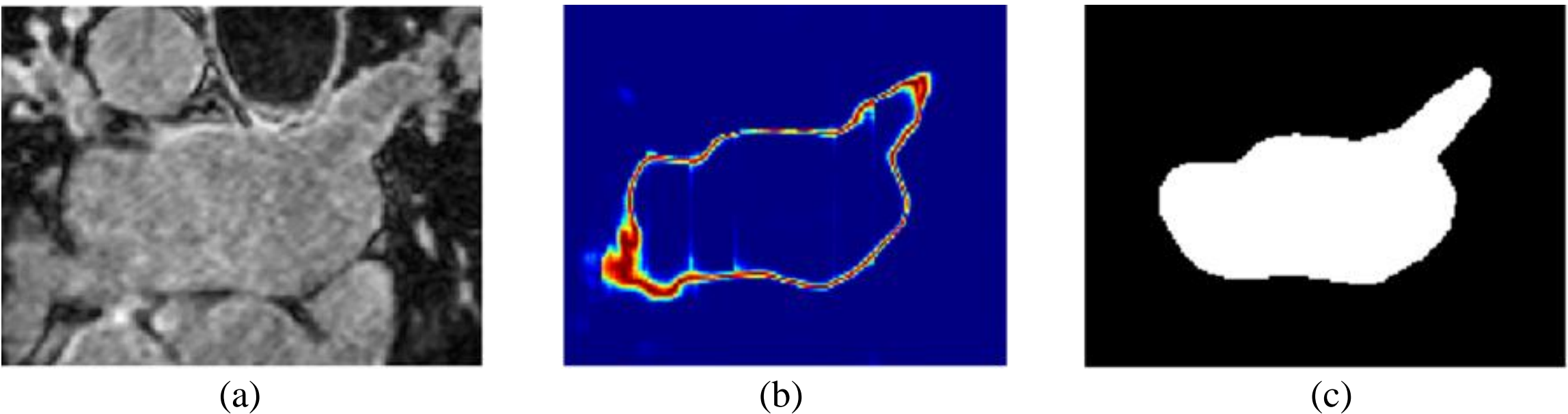

**Fig. 1.** Visualization of the hard-to-segment regions. (a) denotes the original image. (b) denotes the uncertainty of the segmentation results. (c) denotes the corresponding binary ground truth. It can be observed that the regions with high uncertainty mainly located around the object boundary.

The basic idea of self-training is to train an initial model on labeled data to generate pseudo labels for unlabeled data, then the model can be retrained with the updated pseudo-labeled data [15]. Obviously, the qualities of the pseudo-labels determine the performance of this kind of methods. Although several approaches have been proposed to maximally improve the accuracy of the pseudo-labels, incorrect predictions may still exist and be reinforced following this pipeline, especially when annotated data is scarce. The other category is consistency-based methods, which features the smoothness assumption. This paradigm typically adds small perturbations to the input data and leverage unlabeled data through enforcing the prediction consistency between the original data and perturbed data. The representative approaches are self-ensembling methods [16]-[18], like π model [17] and mean-teacher model [18]. Several improved versions have been proposed by designing multiple types of perturbations [19], [20] or adopting the uncertainty estimation [21]-[23].

Although semi-supervised segmentation methods have made great progress in medical image domain, existing methods still have two limitations. The first one is the negligence of disparate importance of different regions for the segmentation task [24]. As shown in Fig.1, the hard-to-segment regions are mainly located round the boundary of the organ or tumor, where contains rich information but has low contrast. Contrastively, the regions without complicated texture information are more likely to be correctly identified. The other limitation is the fact that source images often contain slices with noises and low resolution, which induces high uncertainty prediction [25]. Based on the two observations, the main objective of our work is to pay more attention to the ambiguous boundary regions and perform low uncertainty segmentation prediction.

To address the above issues, we proposed a novel geometry-aware semi-supervised learning framework for medical image segmentation in this paper. Our framework is inspired from the consistency-based strategy, which enforces the prediction consistency of input images under different perturbations. To better exploit the rich information around the ambiguous boundary regions, we introduced an auxiliary prediction task to depict the global geometrical contour of the object and further developed an exponentially weighted consistency loss based on the geometrical shape constraint. In addition, different from previous methods, we performed a dual-view network by applying two decoders with different up-sampling strategies to reduce the prediction uncertainty. The two decoders benefit each other in a cross-supervision way and no extra label information is needed. Experiments indicate that the proposed method can significantly improve the segmentation performance on two public benchmark datasets. Our main contributions are summarized as follows:

(1): A novel geometry-aware semi-supervised learning method is proposed for medical image segmentation, which is simple to implement and has good generalization ability.

(2): An exponentially weighted consistency loss is designed to emphasize hard-to-segment regions under the instruction of geometric boundary constraint. In addition, a dual-view network is proposed to perform low uncertainty prediction and facilitate model training.

(3): Our method shows superior segmentation performance on the public left atrial dataset and brain tumor dataset with only a small amount of labeled data, meanwhile outperforms several state-of-the-art semi-supervised segmentation methods.

## 2. Related Work

In this section, we first review the typical deep learning-based medical image segmentation methods, then recall the literature of semi-supervised segmentation in medical image analysis, which is closely relevant to our work.

### 2.1. Medical Image Segmentation

During the past few years, convolutional neural network has achieved impressive segmentation performance in medical image research community. The mainstream framework is the fully convolutional network (FCN) [26] based encoder-decoder architecture with the representative model like U-Net [2]. The network consists of a symmetrical encoder-decoder path and adopts skip connections to merge the information of different resolutions. Alom et al. [3] presented a recurrent convolutional neural network based on U-Net for medical image segmentation. Liu et al. [4] incorporated graph learning into the segmentation network to jointly exploit local information and long-range dependencies among tumors at different locations, which performed well in ovarian tumor segmentation. Zhou et al. [5] redesigned the skip connections and proposed U-Net++. The decoders were densely connected at the same resolution via the skip connections and deep supervision was applied to improve the overall segmentation performance. Dutande et al. [27] proposed a modified U-Net by applying separable convolutional residual units and ensemble strategy. Yeung et al. [28] designed a novel attention-gated U-Net architecture, which incorporated both spatial and channel-based attention with a focal parameter to control the degree of background suppression.

The encoder-decoder architecture has also been extended to 3D image segmentation, like 3D U-Net [29] and V-Net [6], which adopted 3D convolution kernels for feature extraction. Fabian Isensee et al [30] proposed a more universal segmentation framework named nnU-Net. The preprocessing, network configuration and training was performed automatically based on the experimental rules, which was adaptive to different datasets.

Most of these methods dedicate to investigate better skip connections or feature aggregation modules, which effectively promote the research of automatic segmentation of medical images. However, these methods may suffer the overfitting problem due to the limited labeled data in the clinical practice.

### 2.2. Semi-supervised Medical Image Segmentation

To alleviate the heavy burden of manual delineation, semi-supervised medical image segmentation has been widely studied for a long period. Early approaches mainly relied on hand-crafted features to perform segmentation. You et al. [31] applied radial projection and semi-supervised self-training method to extract the vessel structures from fundus images. Gu et al. [32] constructed forest oriented super pixels for vessel segmentation, which improved the prediction accuracy of the low confidence regions. These semi-supervised methods generally require prior knowledge and are parameter sensitive, which limits the efficiency and accuracy of the segmentation models.

Recent semi-supervised medical image segmentation methods are mainly based on deep learning due to its strong capability to automatically learn high-level feature representations. Bai et al. [33] proposed a self-training scheme for cardiac segmentation, which generated pseudo labels for unlabeled data and updated the network parameter iteratively. Adversarial learning was another way to perform semi-supervised segmentation [34]-[36]. Zhang et al. [34] applied adversarial learning to make unlabeled images produce similar segmentation outputs as labeled images. Fang [35] combined adversarial learning to discriminate whether the pixels were predicted or from the ground truth, which promoted the biomedical image segmentation accuracy. Multiview co-training has also become a popular solution for semi-supervised medical image segmentation [15], [37], [38]. As a basic example, Xia et al. [37] performed different co-training views by rotating or permuting the 3D image data. Furthermore, Bayesian-estimation uncertainty was adopted to produce more reliable pseudo labels.

More recently, there has been increasing interest to perform semi-supervised segmentation through consistency regularization, which leverages unlabeled data by enforcing prediction consistency with different input perturbations. Laine [17] first proposed a simple but efficient scheme called π model, which imposed prediction consistency of the same sample with different noise perturbations. The temporal ensembling strategy extended π model by considering predictions of previous epochs to perform more reliable predictions. However, the temporal ensembling model requires large space for history data storage and the parameters cannot be updated in real time. Lately, the mean-teacher framework [18] was presented to deal with this issue. A teacher model is maintained by exponential moving average the weights of the student model while the student model learns from the teacher model through the prediction consistency. Based on this basic framework, many approaches were proposed and made further improvement by introducing uncertainty estimation and different perturbations of the input [16], [20]-[22]. For example, Yu et al. [16] designed an uncertainty-aware scheme to encourage the student model gradually learn from the regions with lower uncertainty. Li et al. [20] presented a transformation-consistency strategy to enhance the regularization effect for several medical image segmentation tasks. Dual-task consistency was developed by Luo et al. [39], which extended the regularization from data level to task level.

It can be seen that recent methods mainly focus on investigating better perturbations to enhance the feature representation ability but neglect the disparate importance of different regions for the segmentation prediction task. Our framework performs consistency regularization from different views with geometric awareness, aiming to pay more attention to the ambiguous boundary regions to boost the segmentation performance.

## 3. Methodology

The overall structure of the proposed geometry-aware semi-supervised learning framework for medical image segmentation is illustrated in Fig. 2. The network adopts V-Net [6] as the basic backbone, while a parallel decoder is added to increase the model diversity. The geometrical information is explored by an extra prediction task and further used to instruct the model pay more attention to the hard-to-segment regions. In this section, the network architecture will be first introduced. The details of the geometry-aware consistency regularization scheme and the exponentially weighted strategy will be described in order.

### 3.1. Overview of the network architecture

As mentioned above, the network consists of two branches to perform dual-view learning, aiming to conduct segmentation with model perturbation and reduce the prediction uncertainty. The two decoders are identical in structure but with different up-sampling strategies, one employs original deconvolution like V-Net [6] while the other uses tri-linear interpolation. Each branch implements two different tasks for image segmentation and geometrical shape awareness. In order to leverage unlabeled data, an exponentially weighted cross consistency regularization strategy is designed between the predicted segmentation map and the geometrical distance-derived segmentation map from the other decoder.

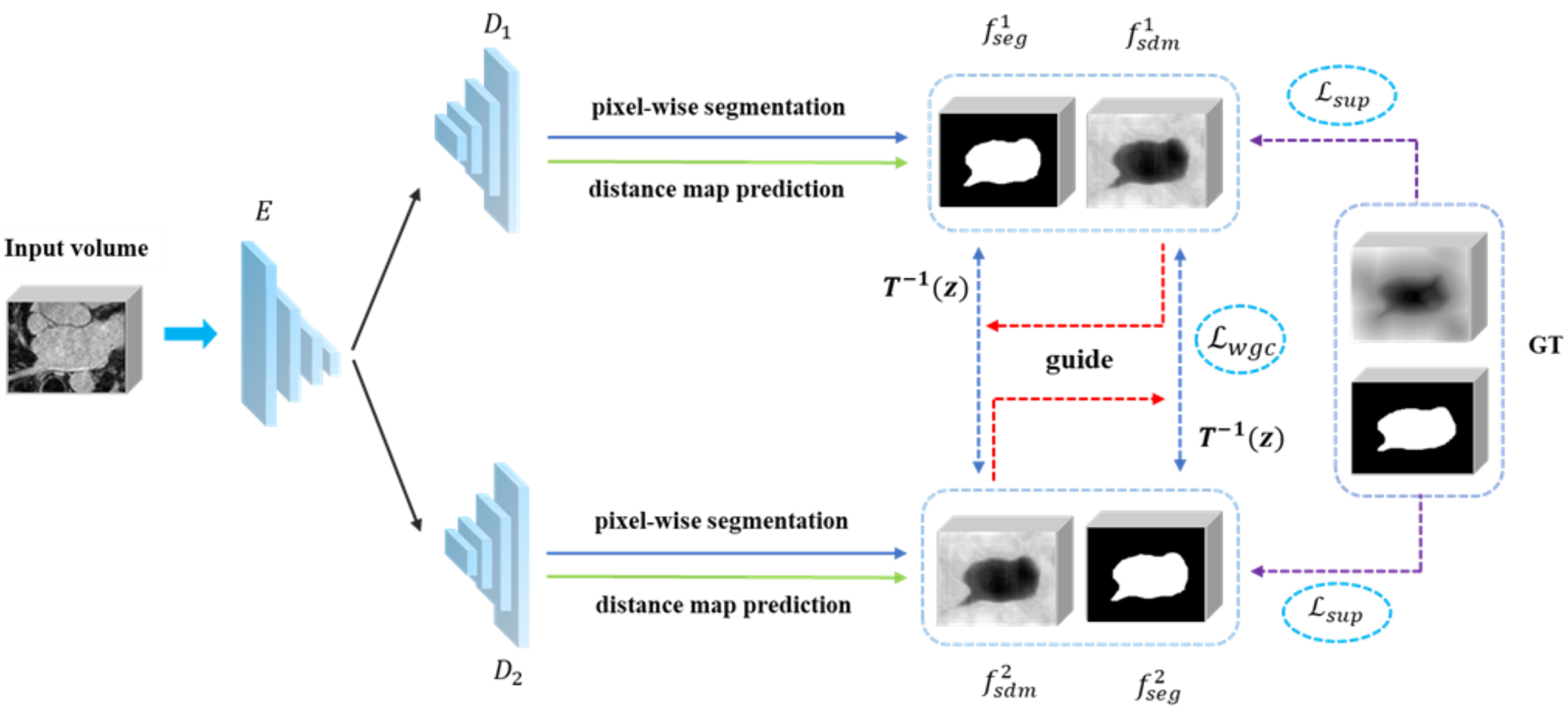

**Fig. 2.** The overview of the proposed geometry-aware semi-supervised segmentation framework. The network consists of a shared encoder $E$ and two decoders $D_1$, $D_2$ with different up-sampling strategies. Each decoder receives the same deep features from the encoder $E$ and jointly predicts a pixel-wise segmentation map $f_{seg}$ and a geometrical-aware signed distance map $f_{sdm}$. The framework leverages unlabeled images through consistency learning between $f_{seg}$ and distance-derived segmentation map $T^{-1}(f_{sdm})$ across the two decoders, with an exponentially weighted strategy based on the geometrical constraint. The model is optimized by minimizing supervised loss $\mathcal{L}_{sup}$ on labeled images and weighted geometry-aware consistency loss $\mathcal{L}_{wgc}$ on both labeled and unlabeled images.

### 3.2. Geometry-aware Consistency Regularization

Consistency loss is usually designed at the data-level under the smooth assumption in general consistency-based semi-supervised segmentation methods. In contrast to previous methods, signed distance map is another way to embed geometrical contours of the object with extra supervised information. It can be easily incorporated with the segmentation network for prediction and no extra label information is needed. The transformation function below shows how to obtain signed distance map from the binary ground truth image.

$$T(x) = \begin{cases} -\inf\limits_{y\in\partial G} \|x-y\|_2 & x \in G_{in} \\ 0 & x \in \partial G \\ +\inf\limits_{y\in\partial G} \|x-y\|_2 & x \in G_{out}, \end{cases} \quad (1)$$

where $\|x-y\|_2$ is the Euclidian distance between voxel $x$ and $y$, $y$ is the voxel closest to $x$ on the object boundary. $G_{in}$, $\partial G$, $G_{out}$ denote the inside, boundary and outside region of the target object, respectively.

Divided by the object contour, the voxel inside the object takes the negative value and otherwise positive. During the forward prediction, the signed distance map can be generated by an extra 3D convolutional block followed by a hyperbolic tangent activation function. Since the signed distance map reflect the object boundary from the geometric space, it should be feasible to map the signed distance map to the binary segmentation map. Here, we adopted a smooth approximation function to realize the conversion followed previous work [39], which can be defined as:

$$T^{-1}(z) = \frac{1}{1+e^{-k\cdot z}} = \sigma(k \cdot z), \quad (2)$$

where $z$ means the distance value of the voxel $x$. $k$ is a hyper-parameter and is selected as large as possible to approximate the segmentation ground truth.

The effect of the approximation transformation is similar to the sharpen function [35], aiming to achieve low-entropy prediction and make contribution to entropy regularization.

The transformed segmentation map is derived from the geometric contour information while the original segmentation map is based on semantic information. Therefore, it is meaningful to establish a consistency regularization to utilize information from the two perspectives. Particularly, since the two different decoders increase diversity of the segmentation models, the consistency regularization is designed across the two decoders to encourage mutual consistency learning. For both labeled and unlabeled data, we define a geometry-aware consistency loss as follows:

$$\mathcal{L}_{gc}(x) = \sum_{x_i \in D} \left( \left\| f_{seg}^1(x_i) - T^{-1}(f_{sdm}^2(x_i)) \right\|^2 + \left\| f_{seg}^2(x_i) - T^{-1}(f_{sdm}^1(x_i)) \right\|^2 \right), \tag{3}$$

where $f_{seg}^1$ and $f_{seg}^2$ represent the segmentation predictions of the two decoders, while $f_{sdm}^1$ and $f_{sdm}^2$ represent signed distance map predictions of the corresponding decoders.

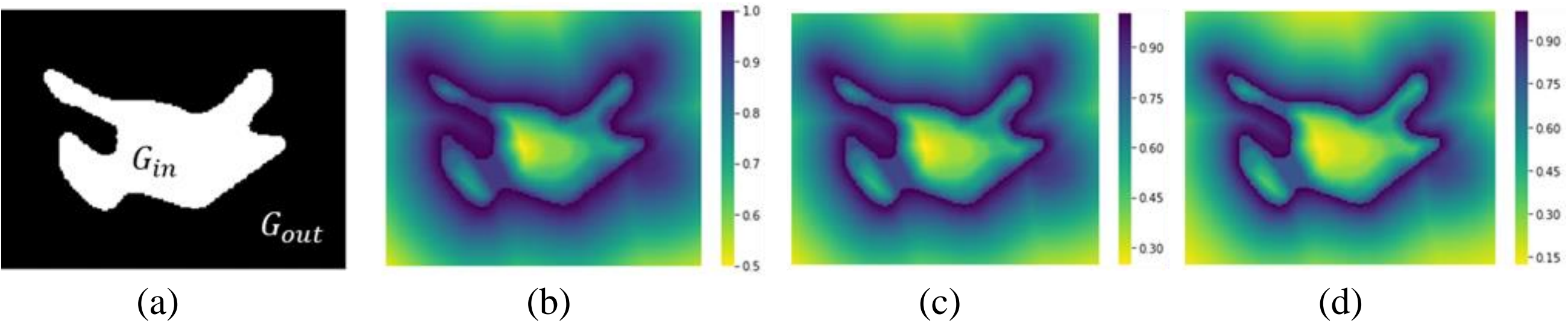

**Fig. 3.** The exponentially weights derived from the geometrical boundary constraint. (a) denotes the binary ground truth image. (b), (c), (d) denotes the weight distribution with $\rho = 1, \rho = 2, \rho = 3$, respectively.

The discrepancy of the two branches can lead to inherent prediction perturbations, which encourages the model to learn better feature representations from different embedding spaces and model diversity.

### 3.3. Exponentially Weighted Strategy

As mentioned above, the awareness of the global geometric information helps the model to better exploit the object structure. However, the segmentation of ambiguous boundary regions still remains a challenge. Intuitively, the boundary regions contain more useful texture information, which is more valuable for the segmentation task. Therefore, we considered to emphasize these hard samples during training by assigning different weights according to their geometric characteristics. The prediction of the signed distance map provides the closest distance to the object boundary of each voxel, which not only brings additional supervision information, but also relatively indicates the segmentation difficulty of different regions. From this point of view, we designed an exponentially weighted strategy to pay more attention to the ambiguous boundary regions. The weights calculation is defined in Eq. (4), where $j$ refers to different decoders and $\rho$ is a hyper-parameter to control the distribution of the weights value. Fig. 3 shows the weight distribution with different selection of $\rho$ . In this way, the voxel near the boundary takes larger weight while the voxel far away from the boundary takes lower weight. Particularly, since we have two different decoders, the weighted strategy also follows a cross supervised way, as Eq. (5) shows.

$$\omega_j = e^{-\rho \cdot |f_{sdm}^{j}(x)|} \quad j = 1, 2 \tag{4}$$

$$\mathcal{L}_{wgc}(x) = \sum_{x_i \in D} \left(\omega_1 \cdot \left\| f_{seg}^{1}(x_i) - T^{-1}(f_{sdm}^{2}(x_i)) \right\|^2 + \omega_2 \cdot \left\| f_{seg}^{2}(x_i) - T^{-1}(f_{sdm}^{1}(x_i)) \right\|^2 \right). \tag{5}$$

### 3.4. Overall Training Pipeline

To better illustrate the overall training pipeline, we first formulate a standard semi-supervised learning setting, in which the training set $D$ contains $N$ labeled images and $M$ unlabeled images, where $N \ll M$ . Let $D^L = \{X_n, Y_n\}_{n=1}^{N}$ denotes labeled dataset and $D^U = \{X_m\}_{m=1}^{M}$ denotes unlabeled dataset. $X$ and $Y$ refer to the input image and the ground truth segmentation mask respectively.

For labeled data $D^L$, we adopt commonly used Dice loss and cross-entropy loss as supervised segmentation loss in Eq. (6). Outputs of the two decoders were both taken into consideration.

$$\mathcal{L}_{seg}(\theta; \theta'; D^L) = 0.5 \times (\mathcal{L}_{dice}(f_{seg}^{1}, y) + \mathcal{L}_{ce}(f_{seg}^{1}, y)) + 0.5 \times (\mathcal{L}_{dice}(f_{seg}^{2}, y) + \mathcal{L}_{ce}(f_{seg}^{2}, y)) \tag{6}$$

As we introduce an auxiliary task for the signed distance map prediction, the prediction error of the signed distance map is also taken into consideration, as Eq. (7) shows. $f_{sdm}$ is the output signed distance map and $T(y)$ is the ground truth transformed from label $y$.

$$\mathcal{L}_{sdf}(\theta;\theta';D^L) = \frac{\left\|f_{sdm}^1 - T(y)\right\|^2 + \left\|f_{sdm}^2 - T(y)\right\|^2}{2} \tag{7}$$

The supervised loss is the weighted sum of the two losses explained above, which can be formulated as Eq. (8). The supervised distance loss helps the model output a reliable distance map and promotes mutual consistency learning. It also enriches the feature representation, making the model more robust.

$$\mathcal{L}_{sup}(\theta;\theta';D^L) = \mathcal{L}_{seg} + \beta \cdot \mathcal{L}_{sdf} \tag{8}$$

Since we define the prediction of the signed distance map is an auxiliary task, $\beta$ is set to 0.3 empirically.

The total loss $\mathcal{L}_{total}$ is the weighted sum of supervised segmentation loss $\mathcal{L}_{sup}$ and the exponentially weighted consistency loss $\mathcal{L}_{wgc}$, as Eq. (9) shows. For both labeled and unlabeled data $D^L$ and $D^U$, we perform geometry-aware consistency regularization to learn useful information under the guidance of geometric shape of the object. The goal of the whole framework is to minimize the weighted sum of $\mathcal{L}_{sup}$ and $\mathcal{L}_{wgc}$.

$$\mathcal{L}_{total} = \mathcal{L}_{sup} + \lambda\mathcal{L}_{wgc} \tag{9}$$

Following previous work [16], [39], we employ a Gaussian ramp-up warming function $\lambda(t) = 0.1 \times e^{-5(1-t/t_{max})}$ to balance the supervised loss and the consistency loss, where $t$ and $t_{max}$ refer to the current training step and maximum training step, respectively. Since the predictions may be less reliable at the early training stage, $\lambda(t)$ can make the training process more smoothly and efficiently. The overall training scheme of the proposed framework is illustrated in Algorithm 1.

---

**Algorithm 1** Training scheme of the proposed geometry-aware semi-supervised segmentation method

---

**Input:** $x_i \in D^L + D^U, y_i \in D^L$

**Output:** parameter $\theta, \theta'$ for trained network

1. $f_{seg}^1$ , $f_{sdm}^1$ represent the predictions of segmentation map and signed distance map of decoder 1, respectively.

2. $f_{seg}^2$ , $f_{sdm}^2$ represent the predictions of segmentation map and signed distance map of decoder 2, respectively.

3. **while** not stopping criterion **do**

4. Sample batch $b = \{(x_i, y_i) \in D^L ,\ x_i \in D^U)\}$

5. Generating SDM ground truth $T(y_i)$ according to Eq. (1)

6. Generating output segmentation maps $f_{seg}^1(x_i)$ , $f_{seg}^2(x_i)$ and signed distance maps $f_{sdm}^1(x_i)$, $f_{sdm}^2(x_i)$

7. Calculating exponential weighted consistency loss $\mathcal{L}_{wgc}$ as Eq. (4)(5)

8. Calculating supervised loss $\mathcal{L}_{sup}$

9. $\mathcal{L}_{total} = \mathcal{L}_{sup} + \lambda \cdot \mathcal{L}_{wgc}$

10. Computing gradient of loss function $\mathcal{L}_{total}$ and update network parameters $\theta, \theta'$

11. **end while**

12. **return** $\theta, \theta'$

---

# 4. Experiments

We have evaluated the proposed geometry-aware semi-supervised segmentation method on two public benchmark datasets: left atrium dataset and brain tumor dataset, with extensive ablation studies and comparison with several state-of-the-art methods. The details of the experiment implementation and the experimental results will be introduced in this section.

### 4.1. Dataset and Pre-processing

The first dataset is the public left atrial dataset [40] from 2018 Atrial Segmentation Challenge. The dataset consists of 100 3D gadolinium-enhanced MR images, with an isotropic resolution of 0.625×0.625×0.625mm. Followed by existing studies[16], [36], [39], 80 scans were used for training and 20 scans were used for testing. The pre-processing strategies are also the same to ensure a fair comparison. In this study, the performance of our method training with 20% labeled images or 10% labeled images are reported, which are the typical semi-supervised segmentation experimental settings.

The second dataset is from the Brain Tumor Segmentation (BraTS) 2019 challenge [43]. The dataset contains multi-institutional preoperative MRI of 335 glioma patients, including 259 high-grade glioma (HGG) patients and 76 low-grade glioma (LGG) patients. Each patient has four modalities of MRI scans including T1, T1Gd, T2 and T2-FLAIR, with neuroradiologist-examined pixel-wise labels. We preprocess the data by means of resampling to $1\,mm^3$ resolution and intensity normalized to zero mean and unit variance. In this study, the dataset was spilt into 250 scans for training, 25 scans for validation and the remaining 60 scans for testing.

### 4.2. Implementation Details and Evaluation Metrics

The whole framework was implemented in Python with PyTorch and trained on a NVIDIA RTX 2080Ti GPU with 11GB memory. We randomly cropped 3D sub-volumes of size 112×112×80 (Left Atrium Dataset) and 96×96×96 (Brain Tumor Dataset) as the model input and applied standard data augmentation like rotation and flip operations. The batch size was set to 4 and each batch contains two labeled images and two unlabeled images. The network was trained by SGD optimizer for 6K iterations, with an initial learning rate 0.01 decayed by 0.1 every 2500 iterations. The hyper-parameter $\rho$ was set to 2.0 in this work.

**Table 1.** Quantitative comparison between the proposed method and other semi-supervised methods on the left atrium dataset using 20% labeled data for training. The first and second rows are the fully supervised baseline.

| Methods | Scans used | | Metrics | | | |
|---|---|---|---|---|---|---|
| | Labeled | Unlabeled | Dice(%) ↑ | Jaccard(%) ↑ | ASD(voxel) ↓ | 95HD(voxel) ↓ |
| V-Net | 80 | 0 | 91.14 | 83.82 | 1.52 | 5.75 |
| V-Net | 16 | 0 | 86.03 | 76.06 | 3.51 | 14.26 |
| DAP[41] | 16 | 64 | 87.89 | 78.72 | 2.74 | 9.29 |
| UA-MT[16] | 16 | 64 | 88.88 | 80.21 | 2.26 | 7.32 |
| SASSNet[36] | 16 | 64 | 89.54 | 81.24 | 2.20 | 8.24 |
| DUWM[21] | 16 | 64 | 89.65 | 81.35 | 2.03 | 7.04 |
| LG-ER-MT[42] | 16 | 64 | 89.62 | 81.31 | 2.06 | 7.16 |
| DTC[39] | 16 | 64 | 89.42 | 80.98 | 2.10 | 7.32 |
| **Ours** | 16 | 64 | **90.34** | **82.49** | **1.70** | **6.57** |

**Table 2.** Quantitative comparison between the proposed method and other semi-supervised methods on the left atrium dataset using 10% labeled data for training. The first and second rows are the fully supervised baseline.

| Methods | Scans used | | Metrics | | | |
|---|---|---|---|---|---|---|
| | Labeled | Unlabeled | Dice(%) ↑ | Jaccard(%) ↑ | ASD(voxel) ↓ | 95HD(voxel) ↓ |
| V-Net | 80 | 0 | 91.14 | 83.82 | 1.52 | 5.75 |
| V-Net | 8 | 0 | 79.99 | 68.12 | 5.48 | 21.11 |
| DAP[41] | 8 | 72 | 81.89 | 71.23 | 3.80 | 15.81 |
| UA-MT[16] | 8 | 72 | 84.25 | 73.48 | 3.36 | 13.84 |
| SASSNet[36] | 8 | 72 | 87.32 | 77.72 | 2.55 | 9.62 |
| DUWM[21] | 8 | 72 | 85.91 | 75.75 | 3.31 | 12.67 |
| LG-ER-MT[42] | 8 | 72 | 85.54 | 75.12 | 3.77 | 13.29 |
| DTC[39] | 8 | 72 | 86.57 | 76.55 | 3.74 | 14.47 |
| **Ours** | 8 | 72 | **88.66** | **79.80** | **1.95** | **7.71** |

During the inference stage, we employed a sliding window strategy with a stride of 18×18×4 for left atrium data and 64×64×64 for brain tumor data to obtain the final segmentation results. We use the output of the decoder with deconvolutional up-sampling strategy as the final segmentation results.

Four frequently-used metrics were adopted to quantitatively evaluate the segmentation results, i.e. Dice similarity coefficient (Dice), Jaccard Index (Jaccard), Average surface distance (ASD) and 95% Hausdorff Distance (95HD). These metrics are complementary to reflect the segmentation performance.

### 4.3. Experiments on Left Atrium Segmentation Dataset

We compare the proposed method with six competitive semi-supervised segmentation methods, including deep adversarial segmentation methods [41], uncertainty-aware mean teacher model [16], shape-aware adversarial network [36], double-uncertainty weighted method [21], local and global structure-aware entropy regularization mean teacher model [42] and dual-task consistency method [39]. Table 1 shows the quantitative comparison results using 20% labeled data for training. As a reference, the segmentation results using V-Net with 16 images (20% labeled data) and all labeled data for training under the fully supervised setting can be viewed as the lower bound and upper bound of the experiment respectively. It can be observed that all seven semi-supervised methods advance the segmentation performance, which proves the effectiveness of exploiting unlabeled data information to boost model training. Compared with other semi-supervised segmentation methods, the proposed method achieves better segmentation performance on all evaluation metrics. The dice and Jaccard coefficient have been improved significantly from 86.03% to 90.34% and 76.06% to 82.49%, which demonstrate the effectiveness of the proposed method. The experiment with 10% labeled data for training shows similar results and are listed in Table 2. The proposed method achieves the best performance among all methods. Notably, some methods encounter significant performance degradation when using less images for training, like

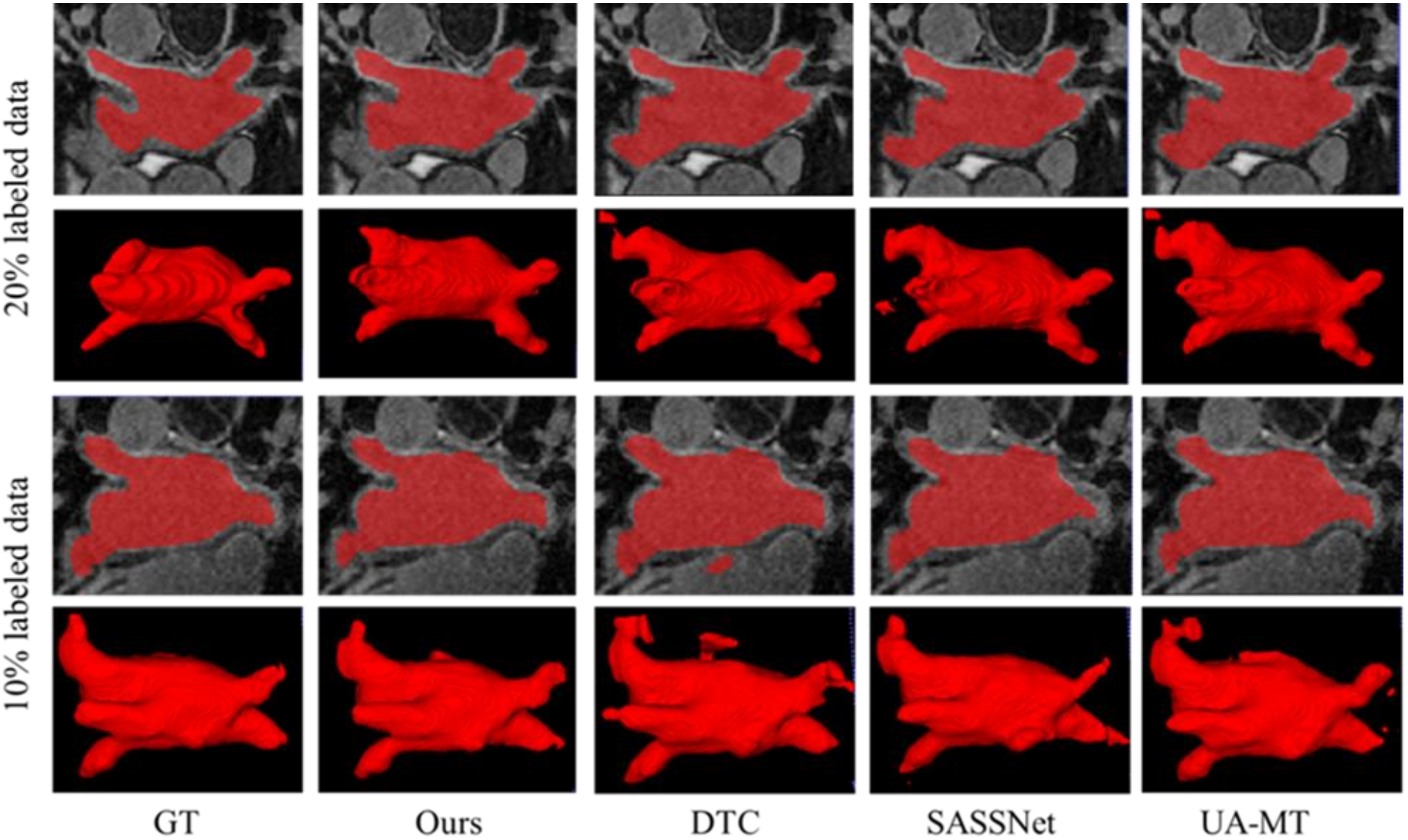

**Fig. 4.** 2D and 3D visualization of different semi-supervised segmentation methods with 20% or 10% labeled data for training. The first column shows the ground truth of the two cases, while other column denotes segmentation results obtained by our methods, DTC [39], SASSNet [36] and UA-MT[16], respectively.

[16], [21], [41], [42]. Contrastively, our method maintains high segmentation accuracy, with Dice coefficient rising from 79.99% to 88.66% and Jaccard coefficient from 68.12% to 79.80%. 2D and 3D Visual comparison of the segmentation results using the proposed method, DTC [39], SASSNet [36] and UA-MT [16] are shown in Fig. 4. It can be observed that the comparison methods tend to wrongly identify some prominent branches, which leads to a large deviation with the ground truth. However, benefiting from the geometrical prior and the exponentially weighted strategy, our method generates better segmentation results at challenging regions with boundary details depicted more accurately. In addition, our method achieves excellent results on the average surface distance (1.70mm/1.95mm) and 95% Hausdorff distance (6.57mm/7.71mm), two boundary-based metrics, which greatly surpass the best score among other semi-supervised methods (i.e., 1.95mm vs

2.55mm in ASD and 7.71mm vs 9.62mm in 95HD). This observation validates the effectiveness of leveraging global geometrical constraint to assist the segmentation process, which leads to more accurate object boundary.

### 4.4. Experiments on Brain Tumor Segmentation Dataset

To further validate the effectiveness of the proposed method, we also conduct experiments on brain tumor segmentation dataset with comparison methods including mean teacher [18], uncertainty-aware mean teacher model [16], dual-task consistency methods [39], shape-aware adversarial network [36] and uncertainty rectified pyramid consistency method [22]. Table 3 presents the segmentation performance under 10% labeled data setting. It can be seen that segmenting brain tumor is a more challenging task. Different from organ segmentation, which owns a relative fixed geometric structure, brain tumors are more diverse in appearance and size. Therefore, leveraging unlabeled data to enrich image information is a promising solution to improve the segmentation performance and alleviate over-fitting problem. As shown in Table 3, all semi-supervised methods achieve performance gain over supervised-only baseline. Furthermore, the proposed method outperforms other state-of-the-art methods on all evaluation metrics. The dice and Jaccard coefficient have been improved significantly from 74.43% to 83.54% and 61.86% to 73.35%, which demonstrate the superiority segmentation performance and good generalization ability of our method. Fig 5 shows visual segmentation results of two brain tumor cases. We can observe that our method generates more accurate segmentation results with less incorrect segmentation regions.

**Table 3.** Quantitative comparison between the proposed method and other semi-supervised methods on the brain tumor dataset using 10% labeled data for training. The first and second rows are the fully supervised baseline.

| Methods | Scans used | | Metrics | | | |
|---|---|---|---|---|---|---|
| | Labeled | Unlabeled | Dice(%) ↑ | Jaccard(%) ↑ | ASD(voxel) ↓ | 95HD(voxel) ↓ |
| V-Net | 250 | 0 | 83.84 | 74.79 | 2.13 | 8.32 |
| V-Net | 25 | 0 | 74.43 | 61.86 | 2.79 | 37.11 |
| MT[18] | 25 | 225 | 81.21 | 70.83 | 2.45 | 14.72 |
| UA-MT[16] | 25 | 225 | 80.85 | 70.32 | 2.57 | 14.61 |
| SASSNet[36] | 25 | 225 | 81.60 | 71.20 | 4.94 | 15.83 |
| DTC[39] | 25 | 225 | 81.96 | 71.84 | 2.43 | 12.08 |
| URPC[22] | 25 | 225 | 81.80 | 71.63 | 2.48 | 11.50 |
| **Ours** | 25 | 225 | **83.54** | **73.35** | **2.13** | **10.47** |

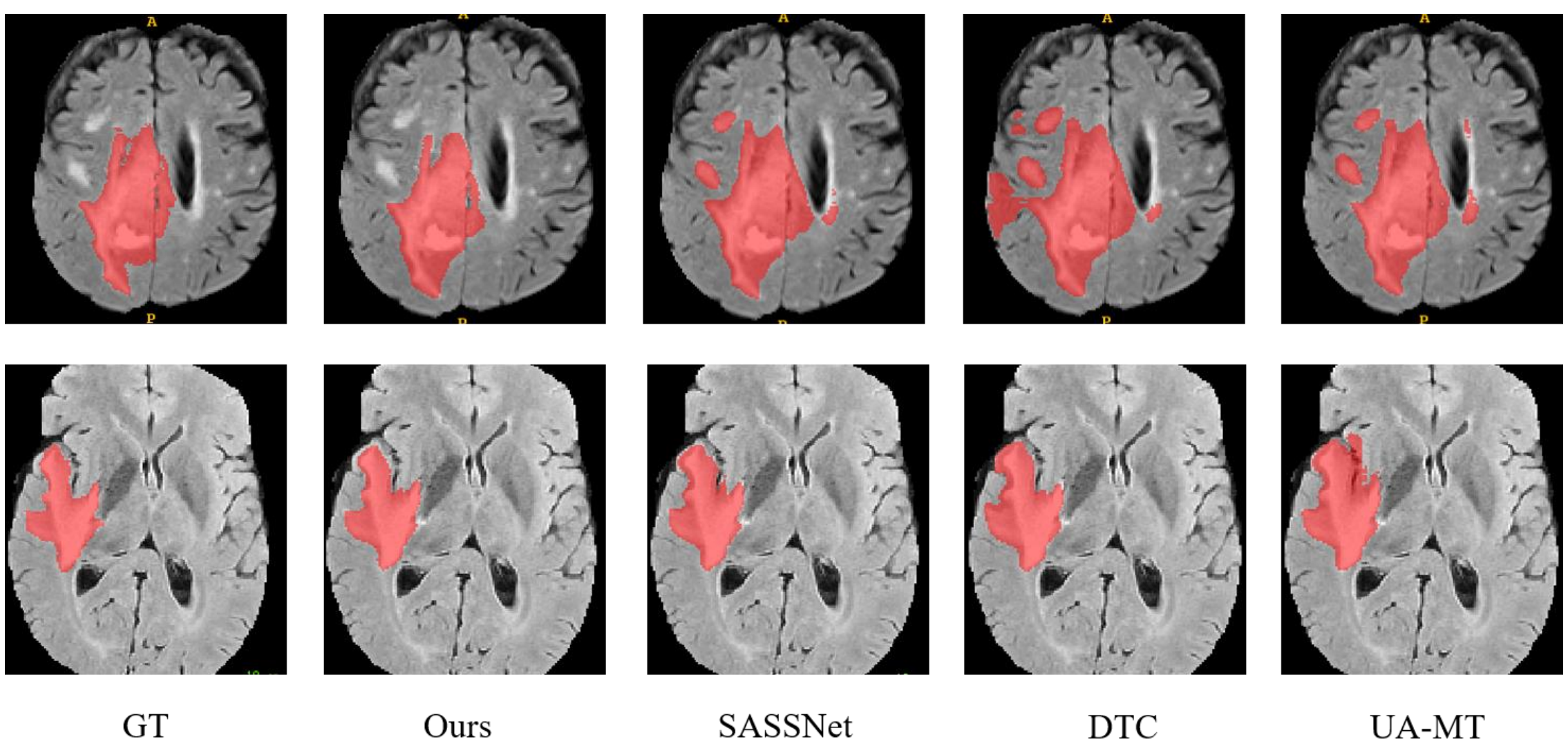

**Fig. 5.** Examples of the brain tumor segmentation results of different semi-supervised segmentation methods with 10% labeled data for training. The first column shows the ground truth of the two cases, while other column denotes segmentation results obtained by our methods, SASSNet [36], DTC [39] and UA-MT[16], respectively.

### 4.5. The Impact of Different Parameter ρ

In order to better explore the rich texture information of the challenging regions around the object boundary, an exponentially weighted strategy is designed to facilitate network training. The hyper-parameter ρ controls the range of the weight value. We investigated the best parameter selection by conducting a series experiments with different value of ρ from 1 to 3. The quantitative segmentation results using left atrium dataset with 10% labeled data were shown in Table 4. It can be found that all five selections improve the segmentation performance compared with the state-of-the-art methods, which demonstrates that paying attention to ambiguous regions is of great help to get better segmentation results. The model achieved the best performance when ρ=2. In fact, the model maintains a high-level performance while the value of ρ between 1.5 and 2.5. A large or small value of ρ may cause slight performance degradation.

**Table4.** Experimental results of different hyper-parameter $\rho$, which controls the range of the weight value.

| $\rho$ | Metrics | | | |
|---|---|---|---|---|
| | **Dice(%)↑** | **Jaccard(%)↑** | **ASD(voxel)↓** | **95HD(voxel)↓** |
| **1.0** | 88.11 | 78.92 | 1.95 | 8.68 |
| **1.5** | 88.50 | 79.52 | 2.04 | 8.48 |
| **2.0** | **88.66** | **79.80** | 1.95 | **7.71** |
| **2.5** | 88.44 | 79.45 | **1.86** | 8.49 |
| **3.0** | 88.27 | 79.19 | 1.94 | 8.64 |

**Table 5.** Effectiveness of the proposed modules on the left atrium dataset with 20% labeled data for training.

| No. | Sup Loss | Cons Loss | Metrics | | | |
|---|---|---|---|---|---|---|
| | | | Dice(%)↑ | Jaccard(%)↑ | ASD(voxel)↓ | 95HD(voxel)↓ |
| 1 | $\mathcal{L}_{seg}$ | - | 88.89 | 80.28 | 1.97 | 8.10 |
| 2 | $\mathcal{L}_{seg} + \mathcal{L}_{sdf}$ | - | 89.25 | 80.74 | 1.89 | 7.57 |
| 3 | $\mathcal{L}_{seg} + \mathcal{L}_{sdf}$ | $\mathcal{L}_{mc}$ | 89.81 | 81.62 | 1.73 | 6.97 |
| 4 | $\mathcal{L}_{seg} + \mathcal{L}_{sdf}$ | $\mathcal{L}_{gc}$ | 89.94 | 81.84 | 1.77 | 6.81 |
| 5 | $\mathcal{L}_{seg} + \mathcal{L}_{sdf}$ | $\mathcal{L}_{wgc}$ | **90.34** | **82.49** | **1.70** | **6.57** |

### 4.6. Ablation Studies

Our framework leverages unlabeled data information by performing geometry-aware consistency learning across two paralleled decoders. To investigate the effectiveness of this design, we performed a series of ablation studies with different constructions of consistency regularization. The segmentation results using 20% labeled data for training are reported in Table 5. Firstly, we used labeled data only for training with different supervised loss. Compared with the fully supervised V-Net trained with 16 labeled images, adopting two different decoders generated better segmentation performance. This shows the dual view training can reduce the prediction uncertainty and lead to more accurate results. The performance has been further improved by adding the signed distance map prediction task to realize geometric awareness. Secondly, we conducted different settings to encourage consistent predictions, i.e. minimizing the discrepancy between the same task predictions of the two decoders (denote as "$\mathcal{L}_{mc}$"), minimizing the discrepancy between different task predictions of the two decoders with or without weighted strategy (denote as "$\mathcal{L}_{gc}$" or "$\mathcal{L}_{wgc}$"). Apparently, encouraging consistency learning to learn from unlabeled images improves the segmentation performance. Moreover, encouraging the consistency predictions achieves better performance when predictions are between different tasks rather than of the same task. That is, performing segmentation from different perspectives enhances the generalization ability and makes the model more robust. Finally, the model achieves the best performance with the exponentially strategy added to instruct the model to pay more attention to the ambiguous regions and boost the learning efficiency.

## 5. Conclusion

In this work, we study the semi-supervised medical image segmentation problem to reduce the human effort of delineating medical image data. A novel semi-supervised learning framework is developed by considering the global geometric information of the object. Our framework can better leverage unlabeled data and generate more accurate segmentation results, which benefits from the dual-view learning and the exploration of ambiguous boundary regions. Extensive experiments on two public benchmark datasets demonstrate the effectiveness of the proposed method. Moreover, the proposed geometry-aware semi-supervised segmentation framework is a general method, which is feasible to be applied to other medical image datasets.

## Declaration of Competing Interest

The authors have no conflict of interest

## Acknowledgements

This work is supported by The National Science Fund for Distinguished Young Scholars (No. 62125306) and the Research Project of the State Key Laboratory of Industrial Control Technology, Zhejiang University, China (ICT2021A15).